\documentclass[referee,sn-nature]{sn-jnl}

\usepackage{graphicx}%
\usepackage{multirow}%
\usepackage{amsmath,amssymb,amsfonts}%
\usepackage{amsthm}%
\usepackage{mathrsfs}%
\usepackage[title]{appendix}%
\usepackage{xcolor}%
\usepackage{textcomp}%
\usepackage{manyfoot}%
\usepackage{booktabs}%
\usepackage{algorithm}%
\usepackage{algorithmicx}%
\usepackage{algpseudocode}%
\usepackage{listings}%

\usepackage{siunitx}
\usepackage{glossaries}
\usepackage{booktabs}
\usepackage{tabularx}

\theoremstyle{thmstyleone}%
\theoremstyle{thmstyletwo}%

\theoremstyle{thmstylethree}%

\DeclareSIUnit\angstrom{\text {Å}}
\DeclareSIUnit\angstromcube{\text {Å$^3$}}

\DeclareRobustCommand{\bm}[1]{\boldsymbol{#1}}

\setacronymstyle{long-short}
\newacronym{dft}{DFT}{density functional theory}
\newacronym{htp}{HTP}{high-throughput}
\newacronym{oqmd}{OQMD}{Open Quantum Materials Database}
\newacronym{vasp}{VASP}{Vienna \textit{ab initio} Simulation Package}
\newacronym{sp}{SP}{spin polarization}
\newacronym{tc}{$T_{{\rm c}}$}{magnetic critical temperature}
\newacronym{mpf}{$\omega_{\mathrm{min}}$}{minimum phonon frequency}
\newacronym{lmm}{$\{\bm{m}_{i}\}$}{local magnetic moment}
\newacronym{mae}{$E_{\mathrm{aniso}}$}{magnetic anisotropy energy}
\newacronym{bz}{BZ}{Brillouin zone}
\newacronym{pmg}{pymatgen}{Python Materials Genomics}
\newacronym{dos}{DOS}{density of states}
\newacronym{ase}{ASE}{Atomic Simulation Environment}
\newacronym{ifc}{IFC}{interatomic force constant}
\newacronym{asa}{ASA}{atomic sphere approximation}
\newacronym{fp}{FP}{full-potential}
\newacronym{icsd}{ICSD}{Inorganic Crystal Structure Database}

\newacronym{ml}{ML}{machine learning}
\newacronym{mlip}{uMLIP}{universal machine learning interatomic potential}
\newacronym{mlrm}{MLRM}{machine learning regression model}
\newacronym{tl}{TL}{transfer learning}
\newacronym{dxmag}{HeuslerDB}{DXMag Computational Heusler Database}
\newacronym{esen}{eSEN-30M-OAM}{}

\newacronym{form}{$\Delta E$}{formation energy}
\newacronym{hull}{$\Delta H$}{energy above the convex hull}
\newacronym{omat24}{OMat24}{Meta Open Materials 2024 Dataset}

\newacronym{hot}{MLIP-HOT}{MLIP High-throughput Optimization and Thermodynamics}
\newacronym{ftl}{MLIP-FTL}{MLIP Frozen Transfer Learning}

\begin{document}

\title{Accurate Screening of Functional Materials with Machine-Learning Potential and Transfer-Learned Regressions: Heusler Alloy Benchmark}


\author*[1]{\fnm{Enda} \sur{Xiao}}\email{Xiao.Enda@nims.go.jp}
\author*[1,2]{\fnm{Terumasa} \sur{Tadano}}\email{Tadano.Terumasa@nims.go.jp}

\affil[1]{\orgdiv{Research Center for Magnetic and Spintronic Materials},
	\orgname{National Institute for Materials Science},
	\orgaddress{\street{1-2-1 Sengen},\city{Tsukuba},
		\postcode{305-0047}, \state{Ibaraki}, \country{Japan}}}

\affil[2]{\orgdiv{Digital Transformation Initiative Center for Magnetic Materials (DXMag)},
	\orgname{National Institute for Materials Science},
	\orgaddress{\street{1-2-1 Sengen}, \city{Tsukuba},
		\postcode{305-0047}, \state{Ibaraki},\country{Japan}}}


\abstract{ We present a machine learning-accelerated high-throughput (HTP)
	workflow for the discovery of functional materials. As a test case,
	quaternary and all-$d$ Heusler compounds were screened for stable compounds
	with large magnetocrystalline anisotropy energy ($E_{\mathrm{aniso}}$).
	Structure optimization and evaluation of formation energy and energy above
	the convex hull were performed using the eSEN-30M-OAM interatomic potential,
	while local magnetic moments, phonon stability, magnetic stability, and
	$E_{\mathrm{aniso}}$ were predicted by eSEN models trained on our DxMag
	Heusler database. A frozen transfer learning strategy was employed to
	improve accuracy. Candidate compounds identified by the ML-HTP workflow were
	validated with density functional theory, confirming high predictive
	precision. We also benchmark the performance of different uMLIPs, discuss the
	fidelity of local magnetic moment prediction, and demonstrate generalization
	to unseen elements via transfer learning from a universal interatomic
	potential. }


\maketitle

\section{Introduction}\label{sec:intro}

The \gls{htp} screening approach has emerged as a powerful strategy for
accelerating the discovery of novel materials by systematically exploring large
chemical spaces computationally or experimentally
\cite{sanvitoAcceleratedDiscoveryNew2017, zhangHighthroughputDesignMagnetic2021,
	barwalLargeMagnetoresistanceHigh2024}. The \gls{dft}-based \gls{htp} workflows
have been widely employed to identify materials with target properties. However,
as the search space increases, the associated computational cost becomes
unsustainable, often restricting screening efforts to a manageable subspace
\cite{faleevHeuslerCompoundsPerpendicular2017,
	huHighthroughputDesignCobased2023}. To address this bottleneck, \gls{ml} offers
a promising route by drastically reducing computational costs. In this work, we
demonstrate the robust integration of state-of-the-art \gls{ml} techniques into
the \gls{htp} workflow (ML-HTP) through a practical case study focused on
screening quaternary and all-$d$ Heusler compounds for stable candidates with
strong \gls{mae}.

Initial realizations of the ML-HTP paradigm relied on \gls{ml} models
that utilize compositional descriptors as input features
\cite{hilgersMachineLearningbasedEstimation2025,
	baigutlinMachineLearningAlgorithms2024, mitraMachineLearningApproach2023,liuMachineLearningPredict2021}. These
models directly map chemical formulas to target properties, offering efficiency
and simplicity. However, composition-based models are inherently unable to
distinguish compounds with identical stoichiometry but different atomic
arrangements. One workaround involves assigning layer indices to atomic sites,
but this approach fixes the number of sites and can yield inconsistent
predictions for symmetry-equivalent
structures~\cite{xieScreeningNewQuaternary2023, luExplainableAttentionCNN2025}.
Crystal graph-based models do not have such drawbacks since they explicitly
incorporate structure information as input, capturing structure-property relationships
more accurately~\cite{xieCrystalGraphConvolutional2018a,
choudharyAtomisticLineGraph2021}. However, crystal graph-based models introduce an additional
computational step, as geometry optimization must precede property prediction.

Although a single \gls{dft} optimization typically requires only a few minutes,
the cumulative cost of screening a large number of candidate compounds becomes
prohibitively expensive, particularly for magnetic systems where multiple
magnetic configurations must be considered. A promising solution lies in
leveraging \glspl{mlip}, which can accelerate structure optimization by several
orders of magnitude relative to \gls{dft}. The \gls{mlip} field has witnessed
rapid advancements in recent years, with many crystal graph-based models
proposed. Despite the conceptual appeal, reliable and robust \gls{mlip}-based
structure optimization has only become practical recently, especially following
the release of the large-scale and diverse \gls{omat24} training dataset in
2024~\cite{barroso-luqueOpenMaterials20242024}. This is demonstrated in the
current work by benchmarking several \glspl{mlip}, ranging from early-stage
implementations to state-of-the-art developments.

With optimized structures, properties can be predicted using \gls{mlrm}. This
approach substantially reduces computational cost, particularly for properties
that are expensive to compute via \gls{dft}, such as phonon spectra,
conductivity, \gls{tc}, and \gls{mae}. However, training accurate \glspl{mlrm}
typically requires large, high-quality datasets. To overcome this challenge,
\gls{tl} techniques can be employed to adapt pretrained \gls{ml} models to new
tasks \cite{yamadaPredictingMaterialsProperties2019,
leeTransferLearningMaterials2021a, hoffmannTransferLearningLarge2023}. \gls{tl}
leverages models that have already learned generalizable representations from
extensive datasets and fine-tunes them using smaller, task-specific datasets.
This strategy enhances predictive accuracy while substantially reducing data
requirements.

As a case study, we conducted a \gls{ml}-\gls{htp} screening on Heusler
compounds, which have garnered significant attention due to their diverse
functional properties, technological potential, and structure complexity
\cite{heComputationallyAcceleratedDiscovery2022}. Numerous DFT-HTP
screenings have been carried out to identify candidates with different desirable
properties~\cite{sanvitoAcceleratedDiscoveryNew2017,
	faleevHeuslerCompoundsPerpendicular2017, nokyGiantAnomalousHall2020,
	huHighthroughputDesignCobased2023,
	xingChemicalsubstitutiondrivenGiantAnomalous2024a,
	xiaoHighthroughputComputationalScreening2025}. In our previous work, we
developed \gls{dxmag}, a comprehensive database encompassing
nearly all conventional ternary Heusler compounds. The present study
significantly extended the search space to include quaternary and all-$d$
Heusler compounds, targeting stability and \gls{mae} as key screening criteria.
In earlier DFT-HTP studies of Heusler compounds, 10 candidates with large \gls{mae} were identified out of
286 selected compositions, and 15 among 29,784 Co-based structures
\cite{faleevHeuslerCompoundsPerpendicular2017,
	huHighthroughputDesignCobased2023}. This low yield underscores the rarity of
such materials and highlights the difficulty of this search problem, making
it a stringent test case for \gls{ml}-\gls{htp} approaches. 

Previous studies attempted ML approaches to $E_{\mathrm{aniso}}$ in systems
outside the Heusler family. They employed early ML models, such as crystal graph
convolutional neural networks (CGCNN) and compositional-descriptor models, to
predict $E_{\mathrm{aniso}}$ in Fe–Co–N alloys and physics-informed 2D
materials, working on the order of hundreds of compounds
\cite{xieDataDrivenStudiesMagnetic2021, liaoPredictingMagneticAnisotropy2022,
duttaMachineLearningAssisted2022}. Here, we employ state-of-the-art ML methods
to extend the scope to hundreds of thousands of compounds with improved accuracy
and practicality.

\begin{figure*}[thbp]
	\centering
	\includegraphics[width=0.99\textwidth]{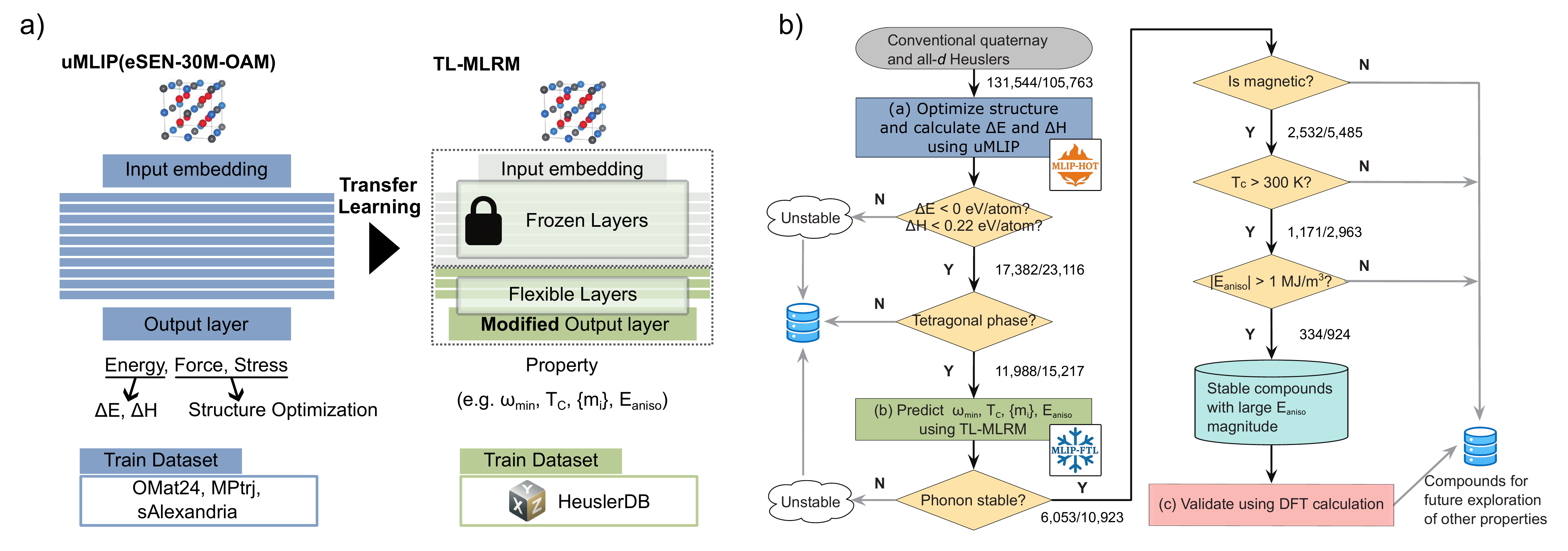}
	\caption{\label{fig:schematic}
	Frozen transfer learning overview and ML-HTP workflow.
	a) Schematic of the development of the \gls{mlrm} via frozen transfer
	learning, using \gls{esen} \gls{mlip} as the base model. The \gls{mlip} is
	used to perform structure optimization, formation energy calculation, and
	convex hull distance evaluation. The \gls{mlrm} predicts properties from
	structures. 
	b) Workflow of the case study which identified stable conventional
	quaternary and all-$d$ Heusler compounds exhibiting strong \gls{mae}. Counts
	of quaternary and all-$d$ compounds at each stage are reported as
	quaternary/all-$d$.}
\end{figure*}

In this work, we demonstrate the use of \gls{mlip} and \gls{tl}-\gls{mlrm}s as
drop-in replacements for \gls{dft} structure optimization and property
evaluation within \gls{htp} framework, as illustrated in
Fig.~\ref{fig:schematic}(a). As a practical application, we employed this
approach to identify conventional quaternary and all-$d$ Heusler compounds with
large \gls{mae}, while simultaneously satisfying thermodynamic, dynamic, and
magnetic stability. Structure optimization and thermodynamic stability
evaluation were performed using the \gls{esen} \gls{mlip} \cite{fuLearningSmoothExpressive2025}. The following
predictions of \gls{lmm}, \gls{mpf}, \gls{tc}, and \gls{mae} were performed
using \gls{mlrm}s. The \gls{mlrm}s were trained via frozen transfer learning,
using \gls{esen} \gls{mlip} as the base model and fine-tuned using \gls{dxmag}
data and newly computed data. ML-selected candidates were validated through
\gls{dft} calculations to demonstrate the significant reliability of this
\gls{ml}-\gls{htp} approach. We further examine key factors that influence the
performance of such ML-HTP workflows, including accuracy of \gls{mlip}-based
structure optimization, the magnetic configuration prediction, the performance of
frozen transfer learning technique and generalization to unseen elements. 
The underlying code for the ML-HTP workflow is made available as open-source
packages \gls{hot} and \gls{ftl}, which can be found on our group's website and
Git repository.

\section{Results}
\label{sec:result}

To accumulate data for training \gls{mae} \gls{mlrm}, we first computed the
\gls{mae} of conventional ternary Heusler compounds within \gls{dxmag} using
\gls{dft}. The \gls{mae} of some Heusler compounds were reported in the previous
work and the agreement between our DFT results and previous work is demonstrated
in Fig.~S1~\cite{faleevHeuslerCompoundsPerpendicular2017,
huHighthroughputDesignCobased2023}. Among all conventional ternary Heusler
compounds, 2190 (7.9 \%) exhibit an \gls{mae} magnitude greater than 1 MJ/m$^3$.
When further screened for thermodynamic, dynamical, and magnetic stability, only
135 compounds (0.5 \%) meet both the high \gls{mae} and stability criteria, 
which are presentated in Table~S2.
These low percentages underscore the difficulty of identifying stable,
high \gls{mae} compounds and highlight the need for more efficient screening
methods demonstrated in current work as a case study.  
The ML-HTP workflow for this case study is summarized in
Fig.~\ref{fig:schematic}(b); detailed computational procedures are provided
in Sec.~\ref{sec:methods}.

For conventional quaternary compounds, we enumerated all combinations where $X$
and $Y$ are transition metals from the $d$-block (excluding Tc and Hg), and $Z$
is a main-group element from groups 13, 14, or 15 of the $p$-block. In addition,
La and Lu were included for $X$ and $Y$ because their 4$f$ orbitals are either
empty or fully filled. This exhaustive enumeration, accounting for symmetry
constraints, yielded 131,544 unique compositions. For the all-$d$ Heuslers, we
extended the screening space to include $d$-block transition metals
together with La and Lu across all four sites ($X_1$, $X_2$, $Y$, and $Z$),
resulting in a separate set of 105,763 unique compositions. A schematic of the
screened chemical space is presented in Supplementary Information as Fig.~S2. 

\subsection{Validation of ML-HTP selected candidates}

\begin{figure*}[htbp]
	\centering
	\includegraphics[width=0.98\textwidth]{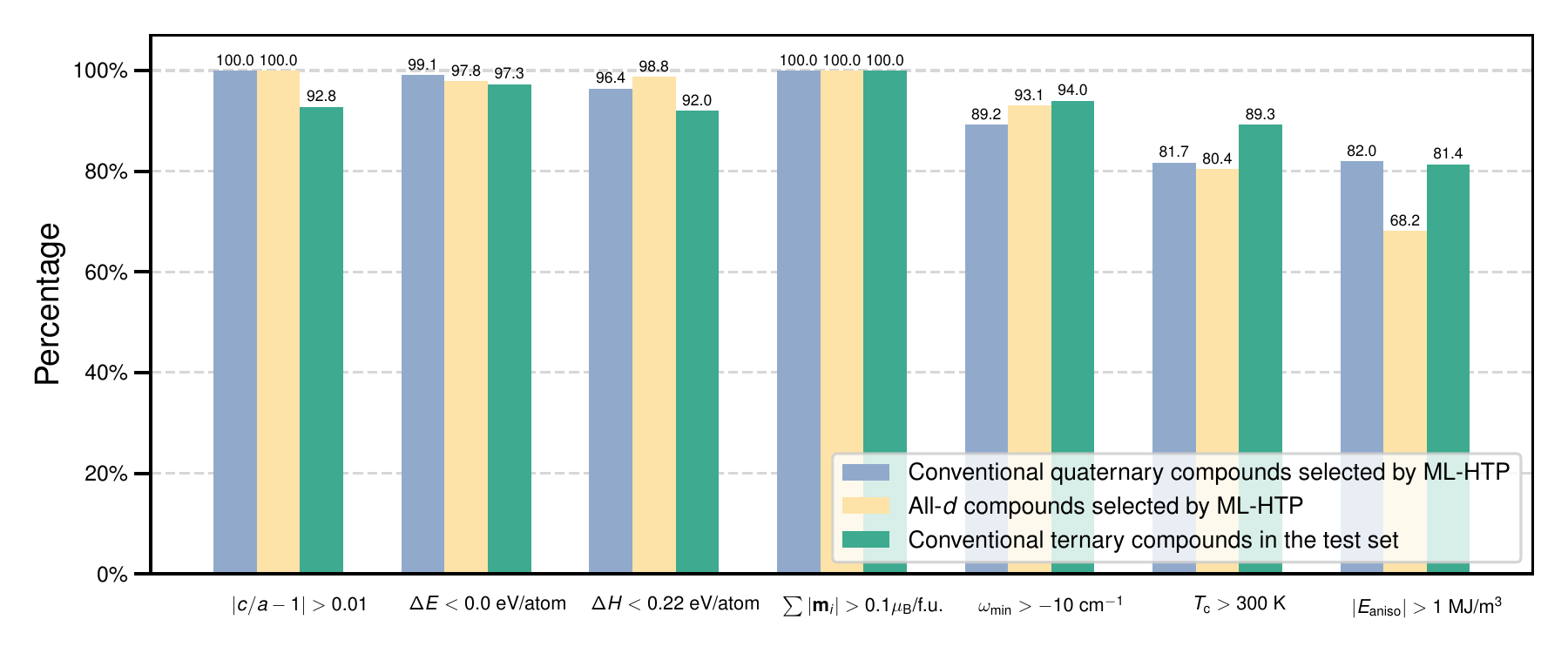}
	\caption{\label{fig:result_sum} DFT validation summary of ML-HTP selected
		compounds. For ML-selected candidate lists of conventional quaternary
		(334) and all-\(d\) (924) Heusler compounds, the percentages that DFT
		results satisfy the screening criteria (i.e., the ML‑HTP precision) are
		shown as blue and yellow bars. For comparison, the precision of the ML
		models measured on the test set of conventional ternary compounds is
		also shown as green bars. The test set size for $c/a$ ratio, \gls{form},
		and \gls{hull} is 10,000 and for \gls{lmm}, \gls{mpf}, \gls{tc}, and
		\gls{mae} is 10\% of the dataset size shown in
		Table~\ref{tab:ml_metrics}.}
\end{figure*}

Using \gls{esen} \gls{mlip} in combination with \gls{mlrm}s, we screened almost
all conventional quaternary and all-$d$ Heusler spaces for stable compounds with
high \gls{mae}. As a result, 334 and 924 candidates were found, respectively. To
evaluate the reliability of this ML workflow, all candidates were validated
using DFT calculations. The results are summarized in Fig.~\ref{fig:result_sum}
and detailed data are provided in Tables~S3 and S4. The percentages that DFT
results satisfy the screening criteria (i.e., precision) are shown as blue and
yellow bars for conventional quaternary and all-$d$ Heusler compounds,
respectively. The selection criteria include $c/a$ ratio, \gls{form},
\gls{hull}, \gls{lmm}, \gls{mpf}, \gls{tc}, and \gls{mae}. For comparison, the
precisions of the ML models, measured on the test set of conventional ternary
compounds, are also provided as green bars. 

In ML-HTP, the $c/a$ ratio, \gls{form}, and \gls{hull} are obtained from
the optimized structure and the corresponding energy by \gls{esen} \gls{mlip}.
During the structure optimization process, relaxations were performed starting
from multiple initial structures, and the relaxed structure with the lowest energy was selected.
Since a non-zero \gls{mae} requires the Heusler compound to adopt a tetragonal
phase, we applied a screening threshold of $|c/a - 1| > 0.01$ to identify
tetragonality. Notably, all ML-selected candidates remained tetragonal in DFT
validation, confirming that \gls{esen} reliably distinguishes between cubic and
tetragonal phases. A more detailed discussion of structure optimization and
performance for lattice parameters $a$, $c$, and the $c/a$ ratio prediction,
along with performance of other \gls{mlip}s, are provided in
Sec.~\ref{sec:mlip_opti}.

Using energies of candidate compounds, elements, and competing phases predicted
by \gls{esen}, the \gls{form} and \gls{hull} were calculated. The criteria of
\gls{form} $< 0$ eV/atom and \gls{hull} $< 0.22$ eV/atom were employed to
identify thermodynamically stable candidates, following the thresholds
established in our previous DFT-HTP study~\cite{xiaoHighthroughputComputationalScreening2025}. Among ML-selected
candidates, 99.1\% of conventional quaternary and 97.8\% of all-$d$ Heusler
compounds were validated to have $\Delta E_{\mathrm{DFT}} < 0$~eV/atom.
Similarly, 96.4\% (quaternary) and 98.8\% (all-$d$) of the compounds were found
to have $\Delta H_{\mathrm{DFT}} < 0.22$~eV/atom. These high validation rates
demonstrate that state-of-the-art \glspl{mlip}, such as \gls{esen}, can reliably
assess thermodynamic stability.

It is important to note that the \gls{esen} \gls{mlip} used is a
general-purpose, pretrained model without any fine-tuning specific to the
Heusler chemical space. Thus, these results highlight its strong generalization,
making it an effective drop-in replacement for DFT-based optimization and
thermodynamic stability assessment in \gls{htp} workflow. Notably, the model
achieves strong performance on the studied magnetic systems, despite not
explicitly incorporating magnetic moments into its architecture or training.
This strong performance and generalization are particularly valuable for the
initial screening of novel material systems, where \glspl{mlip} can greatly
reduce the search space by rapidly and reliably estimating optimized structures
and thermodynamic stability.

The properties \gls{lmm}, \gls{mpf}, \gls{tc}, and \gls{mae} were predicted
using \gls{mlrm}s applied to \gls{mlip}-optimized structures. Because \gls{mae}
is a magnetic property, \gls{lmm} was predicted, and a screening threshold of
$\sum |\bm{m}_{i}| > 0.1$ $\mu_\mathrm{B}$/f.u was applied to identify magnetic
compounds. DFT validations confirmed all ML-selected candidates to be magnetic.
Moreover, the \gls{lmm} \gls{mlrm} accurately predicts both the magnitude and
sign of local moments, as discussed in detail in Sec.~\ref{sec:moment}. Magnetic
system identification is a critical yet computationally demanding step in
DFT-HTP, as multiple initial \gls{lmm} values must be tested and the low
fraction of magnetic systems in some material families can lead to substantial
wasted computation. By incorporating the \gls{lmm} \gls{mlrm} method as a
pre-screening step, the search space can be substantially reduced.

To identify compounds with dynamic stability, magnetic stability, and large
\gls{mae}, we applied the criteria \gls{mpf} $> -10$~cm$^{-1}$, \gls{tc} $>
300$~K, and $|$\gls{mae}$| > 1$~MJ/m$^3$. Among the ML-selected candidates,
89.2\% of conventional quaternary and 93.1\% of all-$d$ Heusler compounds were
validated to have \gls{mpf} above -10~cm$^{-1}$. For magnetic stability, 81.7\%
of conventional quaternary and 80.4\% of all-$d$ Heusler compounds were
validated to have \gls{tc} above 300~K. For the target property \gls{mae}, the
validation rates were 82.0\% and 68.2\%, respectively. To assess the sensitivity
to criteria values, we also evaluated the precisions using a range of more
stringent thresholds. The results, summarized in Fig.~S3, show that the
selection precision is not sensitive to the threshold values in the investigated
range. 

The \gls{mlrm}s were trained exclusively using train set of conventional ternary
Heusler compounds, yet were applied to evaluate quaternary and all-$d$
compositions. By comparing validation rates to the precisions calculated using
the test set of conventional ternary compounds, the \gls{mlrm}s for \gls{lmm} 
and \gls{tc} generalize well to these expanded chemical spaces. In contrast, the
\gls{mae} model exhibits lower performance for all-$d$ compounds. This
discrepancy can be attributed to the difference in chemical environments: while
conventional quaternary compounds retain $Z$-site elements from the
$p$-block—consistent with the training set—all-$d$ compounds introduce $Z$
elements from the $d$-block, which were absent during training. \gls{mae} is a
sensitive property, influenced by subtle details of the electronic structure,
and thus more susceptible to domain shifts than \gls{lmm} or \gls{tc}.

Relaxing the screening thresholds increases the pool of candidate compounds and
might capture promising cases missed initially at the expense of more false
positives. Additionally, the curated list of compounds that meet stability
criteria, and further magnetic system criterion, serves as an efficient starting
point for investigation of other functional properties. For readers
interested in exploring an expanded candidate list, the full set of ML-predicted 
data for 131,544 conventional quaternary and 105,763 all-$d$ Heusler
compounds will be accessible through \gls{dxmag}.

\subsection{Distribution of strong \gls{mae} candidates}

\begin{figure}[htbp]
	\centering
	\includegraphics[width=0.7\textwidth]{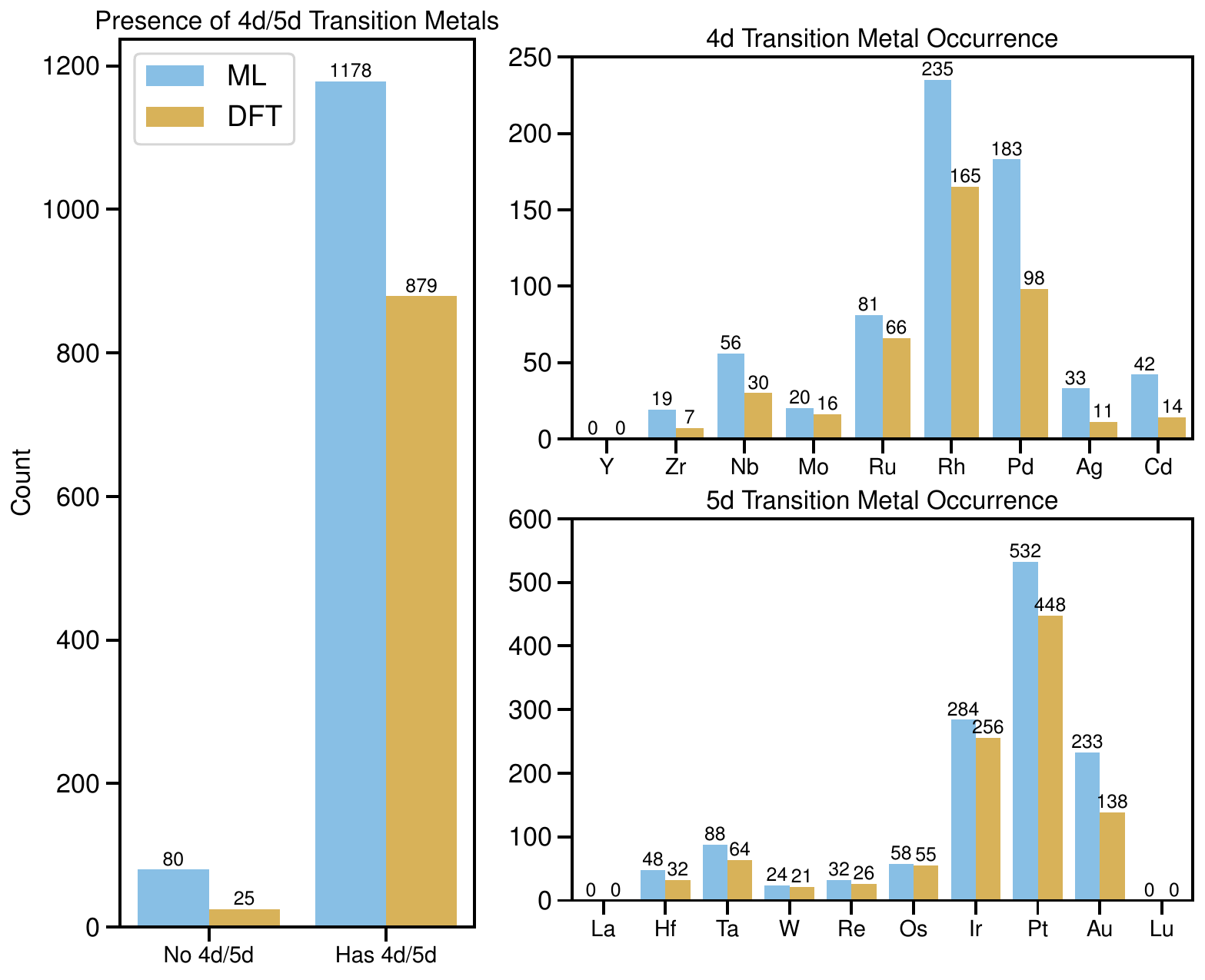}
	\caption{\label{fig:list_distri}  
		Distribution of ML-selected compounds on elements contained.
		Distribution of ML-selected candidate compounds based on whether 4d or
		5d elements are present and distribution over 4d and 5d elements
		contained. The distribution of DFT validated strong \gls{mae} candidates
		is also shown.}
\end{figure}

In addition to validating predictive precision, it is essential to determine
whether the ML models capture known physical trends. A well-established insight
is that compounds containing 4\textit{d} and 5\textit{d} elements typically
exhibit larger \gls{mae} than those composed of 3\textit{d} elements, owing to
the stronger spin--orbit coupling associated with the heavier atomic nuclei of
4\textit{d} and 5\textit{d} elements. This behavior is clearly reproduced in the
ML-HTP results, as shown in Fig.~\ref{fig:list_distri}, which presents the
distribution of candidate compounds according to the presence of
4\textit{d}/5\textit{d} elements. The figure further highlights the specific
4\textit{d}/5\textit{d} elements that appear in the identified compounds. For
comparison, Fig.~\ref{fig:list_distri} also includes the distribution of
compounds with DFT-calculated \gls{mae} magnitudes exceeding 1 MJ/m$^3$.

\subsection{\gls{mlip} optimization performance\label{sec:mlip_opti}}

\begin{figure*}[htbp]
	\centering
	\includegraphics[width=0.85\textwidth]{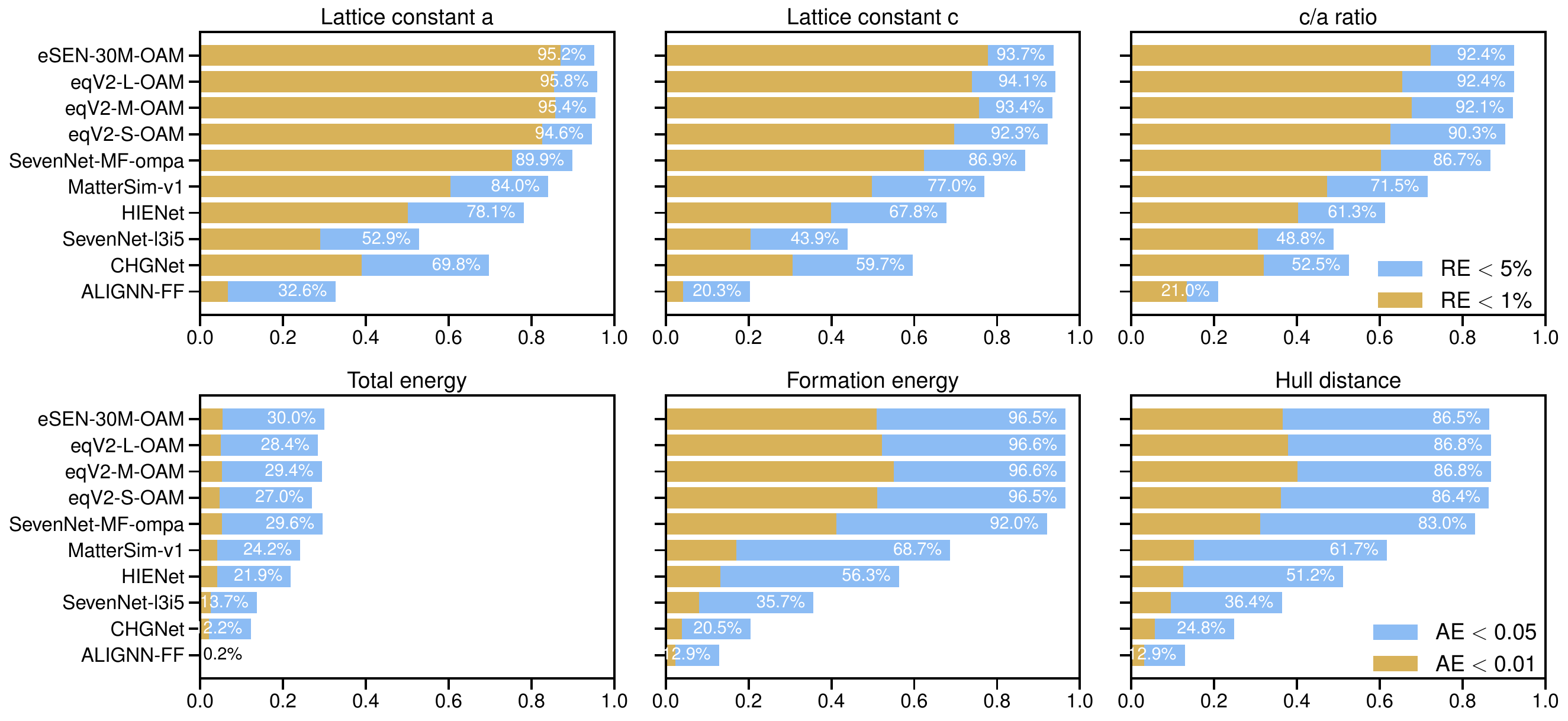}
	\caption{\label{fig:mlip_opti} 
		Benchmark of \gls{mlip} performance.
		Lattice constants $a$ and $c$, $c/a$ ratio, total energy ($E$), formation
		energy (\gls{form}), and convex hull distance (\gls{hull}) predicted by
		various \glspl{mlip} are benchmarked against DFT references. For each
		property, the fraction of compounds with predictions falling within
		specified relative error (RE) or absolute error (AE) thresholds is
		reported. Energetic quantities ($E$, \gls{form}, and \gls{hull}) are
		expressed in eV/atom. The test set consists of 10,000 ground-state
		compounds randomly sampled from \gls{dxmag}.}
\end{figure*}

In recent years, \glspl{mlip} have advanced rapidly, with numerous new models
proposed and trained. To identify the most suitable model for our screening
workflow, we benchmarked representative \glspl{mlip} models developed since
2023. These include ALIGNN-FF, CHGNet, SevenNet-l3i5, SevenNet-MF-ompa, HIENet,
MatterSim-v1, eqV2-S-OAM, eqV2-M-OAM, eqV2-L-OAM, and the latest \gls{esen}
\cite{choudharyUnifiedGraphNeural2023, dengCHGNetPretrainedUniversal2023,
kimDataEfficientMultifidelityTraining2025, yanMaterialsFoundationModel2025,
yangMatterSimDeepLearning2024, barroso-luqueOpenMaterials20242024,
fuLearningSmoothExpressive2025}. The evaluation focused on structure
optimization for 10,000 conventional ternary compounds randomly selected from
the ground states in \gls{dxmag}. To identify the global minimum, 14 initial
structures were generated by applying strain to the conventional cell (two
formula units) and converting it to the primitive cell (one formula unit).
Specifically, the $a$, $b$, and $c$ axes were uniformly scaled by $\pm10\%$ and
$\pm30\%$, or the $c$-axis alone was varied by $\pm10\%$, $\pm20\%$, $\pm30\%$,
$\pm40\%$, and $\pm50\%$. The lowest-energy structure from these relaxations was
selected as the predicted ground state. Convergence tests for all evaluated
models are shown in Fig.~S4, demonstrating that the selected ground states are
well-converged.

The performance of structure optimization was assessed by comparing the
predicted lattice constants $a$ and $c$, and the resulting $c/a$ ratio, with
corresponding \gls{dft} values. The results are summarized in
Fig.~\ref{fig:mlip_opti}. The relative error (RE) is defined as the maximum of
$\left|\frac{x_{\mathrm{ML}}}{x_{\mathrm{DFT}}} - 1\right|$ and
$\left|\frac{x_{\mathrm{DFT}}}{x_{\mathrm{ML}}} - 1\right|$, where
$x_{\mathrm{ML}}$ and $x_{\mathrm{DFT}}$ denote the values predicted by the
\gls{ml} and DFT, respectively. We report the fractions of compounds within 5\%
and 1\% RE tolerances. Among the evaluated models, the \gls{esen} and eqV2
models achieved the highest accuracy at 5\% RE, with \gls{esen} showing slightly
better performance at the stricter 1\% RE threshold. A key distinction between
the two models is the number of local minima encountered: eqV2-L-OAM identified
91,585 local minima, whereas \gls{esen} identified 32,606. Both counts are much
greater than 10,000, confirming the presence of multiple local minima, but are still
significantly less than 10,000×14. This suggests that many different initial
distortions ultimately converge to the same local minimum. Importantly, \gls{esen}
found substantially fewer local minima than eqV2-L-OAM. This difference is
attributed to the smoother energy landscape of \gls{esen}. Convergence tests
also demonstrate that \gls{esen} achieves convergence with fewer initial
structures, resulting in lower computational cost for \gls{htp} purpose.

The predictive performance of total energy ($E$), formation energy (\gls{form}),
and energy above the convex hull (\gls{hull}) using \gls{mlip} was assessed by
comparing the \gls{mlip} predictions with \gls{dft} values, using the absolute
error (AE), $|x_{\mathrm{ML}} - x_{\mathrm{DFT}}|$. The fractions of compounds
with AE below 0.01 and 0.05~eV/atom are shown in Fig.~\ref{fig:mlip_opti}. Among
the benchmarked models, \gls{esen} and the eqV2 variants showed the highest
accuracy for \gls{form} and \gls{hull} at the 0.05~eV/atom threshold, with
\gls{esen} displaying a slight drop in accuracy at the more stringent
0.01~eV/atom level. Predicted total energies from \gls{mlip} were found to be
systematically lower than \gls{dft} values, reducing direct agreement; however,
this offset also applies to elemental references and competing phases, so the
relative quantities \gls{form} and \gls{hull} remain in strong agreement with
\gls{dft}. Given its robust performance in both structure optimization and
thermodynamic stability, \gls{esen} was selected for integration into the ML-HTP
workflow.

\subsection{Improvements Over Existing Approaches\label{sec:ml_metrics}}

Previous studies have reported the performance of composition-based models in
predicting lattice constants, \gls{form}, and \gls{hull} for cubic Heusler
compounds. For comparison, we evaluated metrics of \gls{esen} on the cubic
Heusler subset, with results summarized in Table~\ref{tab:ml_metrics}. The $R^2$
score for the lattice constant $a$ is 0.994, surpassing the previously reported
ranges of 0.80--0.94 across different Heusler types and the values of 0.94,
0.979, and 0.987 in other works~\cite{mitraMachineLearningApproach2023,
huSearchingHighSpin2020, miyazakiMachineLearningBased2021,
luExplainableAttentionCNN2025}. Similarly, the $R^2$ for \gls{form} reaches
0.995, outperforming earlier results of 0.80--0.88 and 0.93,
0.982~\cite{mitraMachineLearningApproach2023, huSearchingHighSpin2020,
luExplainableAttentionCNN2025}. The $R^2$ for \gls{hull} is 0.98, exceeding
prior values of 0.91 and
0.969~\cite{kimMachinelearningacceleratedHighthroughputMaterials2018,
luExplainableAttentionCNN2025}. The root mean squared errors (RMSE) for $a$ and
\gls{form} are 0.023~\AA\ and 0.029~eV/atom, respectively, which are significantly
lower than the 0.11-0.12~\AA\ and 0.117~eV/atom reported in earlier
work \cite{xieScreeningNewQuaternary2023}.

The \gls{mlrm}s used in screening were trained on the data from \gls{dxmag},
supplemented with newly computed \gls{tc} and \gls{mae} values using optimized
structures in \gls{dxmag}. Test set metrics are summarized in
Table~\ref{tab:ml_metrics} and benchmarked against previously reported results.
The \gls{mlrm} for \gls{lmm} achieved an $R^2$ score of 0.989. For comparison
with prior studies that used total magnetization ($m_{\mathrm{total}}$) as the
target property, our model yielded $R^2 = 0.986$ for $m_{\mathrm{total}}$,
exceeding earlier values of 0.75--0.89, 0.82, and
0.927~\cite{mitraMachineLearningApproach2023, liuMachineLearningAssisted2025,
luExplainableAttentionCNN2025}. For \gls{tc}, the model attained $R^2 = 0.91$
and classification accuracy of 0.91, both substantially higher than previously
reported values of $R^2 = 0.76$ and 0.73 and accuracy of
0.73~\cite{hilgersMachineLearningbasedEstimation2025,
hirohataMachineLearningDevelopment2025}.

\begin{table}[h]
	\caption{\label{tab:ml_metrics} Performance comparison of the eSEN
	\gls{mlip} and \gls{mlrm} with ALIGNN models and previously reported
	results. The size of dataset used for \gls{mlrm} is also listed.}
	\begin{tabular}{@{}lccccc@{}}
		\toprule
		Property             & Metric & \gls{esen} & ALIGNN-FF   & Previous reports                                                                               &              \\
		\midrule
		$a$                  & $R^2$  & 0.994      & 0.128       & 0.80--0.94\footnotemark[1], 0.94\footnotemark[2], 0.987\footnotemark[3], 0.979\footnotemark[4] &              \\
		                     & RMSE   & 0.023      & 0.330       & 0.11-0.12\footnotemark[5]                                                                          &              \\
		\gls{form}           & $R^2$  & 0.995      & 0.453       & 0.80--0.88\footnotemark[1], 0.93\footnotemark[2], 0.982\footnotemark[3]                        &              \\
		                     & RMSE   & 0.029      & 0.310       & 0.117\footnotemark[5]                                                                          &              \\
		\gls{hull}           & $R^2$  & 0.980      & 0.330       & 0.91\footnotemark[6], 0.969\footnotemark[3]                                                    &              \\
		\midrule
		Property             & Metric & eSEN MLRM  & ALIGNN MLRM & Previous reports                                                                               & Dataset size \\
		\midrule
		\gls{lmm}            & $R^2$  & 0.989      & —           & —                                                                                              & 27,864       \\
		$m_{\mathrm{total}}$ & $R^2$  & 0.986      & 0.904       & 0.75--0.89\footnotemark[1], 0.82\footnotemark[7], 0.927\footnotemark[3]                         & 27,864       \\
		$\sum|\mathbf{m}_i|$ & $R^2$  & 0.989      & 0.891       & —                                                                                              & 27,864       \\
		\gls{mpf}            & $R^2$  & 0.750      & 0.734       & —                                                                                              & 8,198        \\
		\gls{tc}             & $R^2$  & 0.910      & 0.844       & 0.76\footnotemark[8], 0.73\footnotemark[9]                                                     & 2,106        \\
		                     & Accu.  & 0.910      & —           & 0.73\footnotemark[8]                                                                           &              \\
		\gls{mae}            & $R^2$  & 0.680      & 0.592       & —                                                                                              & 6,123        \\
		\botrule
	\end{tabular}

	\footnotetext{%
		\textsuperscript{1}\cite{mitraMachineLearningApproach2023} Dataset size is about 1000,
		\textsuperscript{2}\cite{huSearchingHighSpin2020} Dataset size is about 65,000, \\
		\textsuperscript{3}\cite{luExplainableAttentionCNN2025} Dataset size is about 500,000 for $a$ and $m_{\mathrm{total}}$, and about 450,000 for \gls{form} and \gls{hull}, \\
		\textsuperscript{4}\cite{miyazakiMachineLearningBased2021} Dataset size is 143,
		\textsuperscript{5}\cite{xieScreeningNewQuaternary2023} Dataset size is 16,272,
		\textsuperscript{6}\cite{kimMachinelearningacceleratedHighthroughputMaterials2018} Dataset size is 426,148, \\
		\textsuperscript{7}\cite{liuMachineLearningAssisted2025} Dataset size is 1153,
		\textsuperscript{8}\cite{hilgersMachineLearningbasedEstimation2025} Dataset size is 408,
		\textsuperscript{9}\cite{hirohataMachineLearningDevelopment2025} Dataset size is 6500.
	}
\end{table}

To the best of our knowledge, no previous study has predicted phonon stability
or \gls{mae} of Heusler compounds using \gls{ml}. The effectiveness of the
\gls{mpf} and \gls{mae} models developed here is supported by the validation
results in Sec.~\ref{sec:result}. While phonon stability could also be assessed
using \gls{mlip} combined with phonon calculation methods, our regressor-based
approach is motivated by both efficiency and accuracy. Only the minimum frequency is
needed for phonon stability assessment, thus a regressor is sufficient and much
faster than calculating the full spectrum. We also evaluated \gls{mlip}+phonon
using CHGNet, MatterSim, and eSEN-30M-OAM on a test set of 1000 conventional
ternary compounds, and found stability/instability classification accuracies of
62.5\%, 74.9\%, and 80.2\%, respectively, which are substantially below the
regressor's 93.6\%. Notably, CHGNet misclassifies 67.2\% of stable compounds as
unstable, showing a strong tendency to underestimate stability, while eSEN
weakly overestimates it and MatterSim exhibits a more balanced performance.
Additional details including example phonon spectra by \gls{mlip}s and DFT are
in the Supplementary Information. The \gls{mae} model achieved an $R^2$ of 0.68,
lower than those of \gls{lmm}, \gls{mpf}, and \gls{tc}, highlighting the higher
sensitivity and complexity of \gls{mae} as a target property. Nevertheless,
despite the reduced $R^2$, its classification accuracy remains satisfactory and
sufficient for integration into the ML-HTP workflow.

To benchmark advances in \gls{ml} techniques since 2023, we applied the
ALIGNN-FF \gls{mlip} and ALIGNN \gls{mlrm} to identify conventional quaternary
candidate compounds and evaluated the validation rates of strong \gls{mae}
compounds \cite{choudharyUnifiedGraphNeural2023,
choudharyAtomisticLineGraph2021}. In this test, the scalar quantity $\sum
|\mathbf{m}_i|$ was used directly as the target property rather than being
calculated from \gls{lmm} prediction. The metrics of ALIGNN-FF and ALIGNN
\gls{mlrm} are summarized in Table~\ref{tab:ml_metrics}. Using all screening
thresholds, only 17 compounds qualified as candidates. To improve statistical
robustness, we removed the phonon stability criterion, expanding the candidate
list to 107 compounds, of which 26 (24.3\%) exhibit $|$\gls{mae}$| > 1
\mathrm{MJ/m^3}$. While this yield is notably higher than the 7.9\% obtained
from direct DFT-HTP screening, it remains far below the 82.0\% success rate
achieved with the eSEN-based ML-HTP workflow.  These results highlight the
substantial improvements in screening precision enabled by the state-of-the-art
eSEN model.

We further tested a hybrid workflow in which structure optimization was
performed with \gls{esen} \gls{mlip}, while property prediction was carried out
using ALIGNN \glspl{mlrm}. This approach yielded 276 candidate compounds, of
which 149 (54.0\%) were confirmed by DFT to exhibit strong \gls{mae}. The
improved yield relative to ALIGNN-FF based optimization underscores the critical
importance of accurate structure optimization with \gls{esen} for enhancing
ML-HTP screening. However, the yield still falls short of the 82.0\% achieved by
the fully eSEN-based workflow, indicating that progress in both the \gls{mlip}
and \gls{mlrm} components is essential for maximal efficiency. We also tested
the inverse hybrid configuration, using ALIGNN-FF \gls{mlip} for structure
optimization combined with eSEN \glspl{mlrm} for property prediction. This
workflow identified 243 candidates, of which only 76 (31.3\%) were validated as
strong \gls{mae} compounds. This marked reduction in performance highlights the
pivotal role of selecting an accurate \gls{mlip} for structure optimization.

\subsection{Prediction of local magnetic moment \label{sec:moment}}

\begin{figure}[htbp]
	\centering
	\includegraphics[width=0.95 \textwidth]{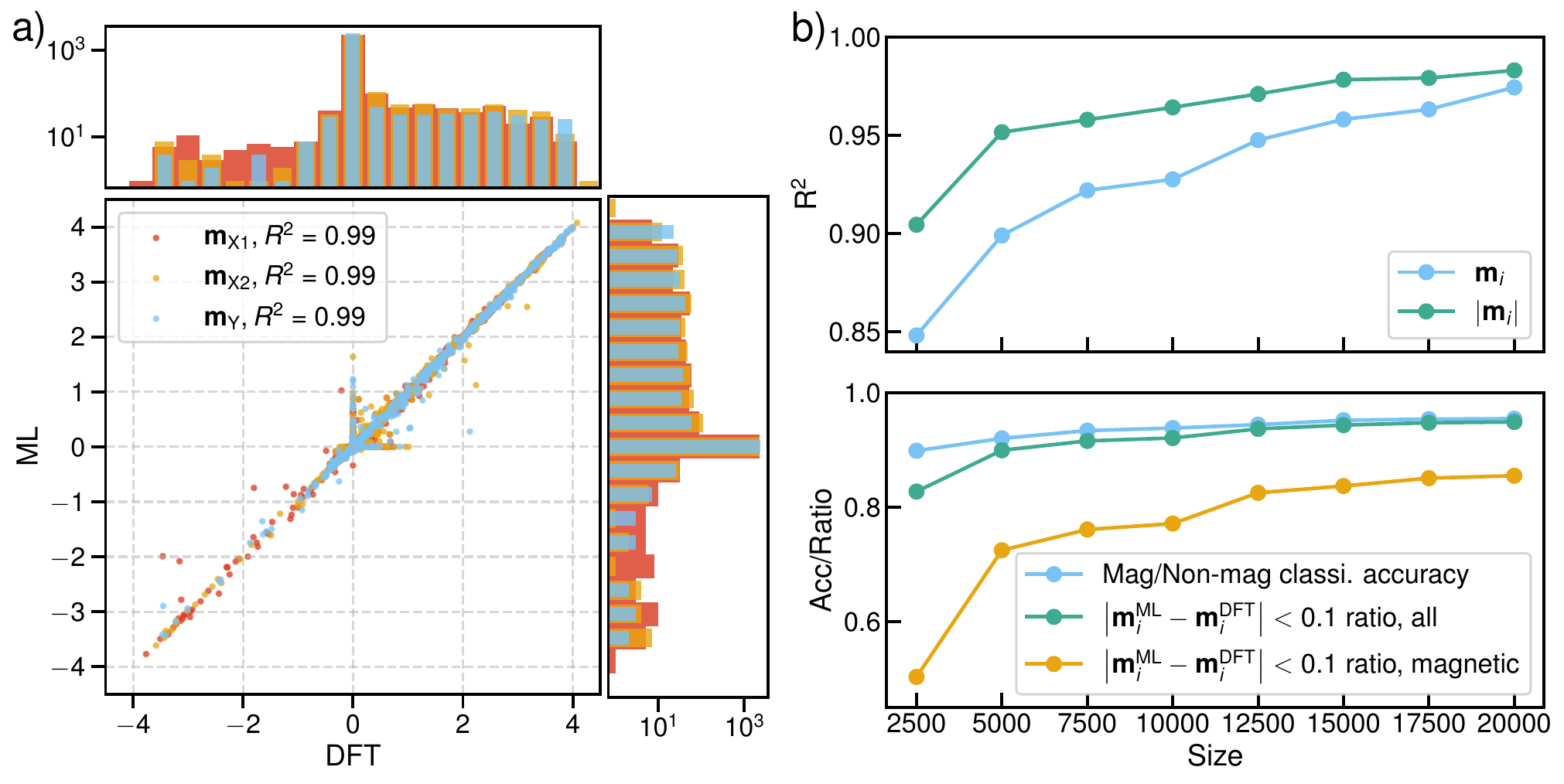}
	\caption{\label{fig:MLRM_mag} 
		Local magnetic moment prediction performance.
	    a) Scatter plot comparing ML-predicted \gls{lmm} with DFT values for
		the test set. 
		b) Learning curves for \gls{lmm} prediction. The top panel shows $R^2$
		scores for both local moments and their magnitudes. The bottom panel
		reports the magnetic/nonmagnetic classification accuracy, and the
		fraction of compounds with absolute prediction error below
		0.1~$\mu_{\mathrm{B}}$ for all compounds and for the magnetic subset.}
\end{figure}

Since the goal of this study is to identify compounds with large \gls{mae}, it
is first necessary to determine whether a compound is magnetic. Relying solely
on the total magnetization is inadequate, as it cannot capture antiferromagnetic
(AFM) or low-moment ferrimagnetic (FiM) compounds. To address this, we employed
the total absolute magnetic moment, defined as $\sum_{i} |m_{i}|$, where
$m_{i}$ denotes the local magnetic moment at atomic site $i$.

Local magnetic moments were predicted using an \gls{mlrm} based on the eSEN
architecture, trained to output the moment at each atomic site. Restricting to
collinear configurations in which all moments are aligned along the $z$-axis, each
moment is represented by a scalar whose sign encodes the direction, with
$\ell=0$ per site in the output head. To account for the $z$-direction
ambiguity—where a configuration and its sign-inverted counterpart (i.e., all
local moments flipped) are physically equivalent—we modified the loss function
to compute losses for both the predicted \gls{lmm} and its sign-inverted
counterpart, and take the smaller value as the loss. This ensures invariance
under global spin inversion.

Fig.~\ref{fig:MLRM_mag} (a) shows a scatter plot comparing \gls{lmm} from the
\gls{mlrm} and \gls{dft} for compounds in test set, with histograms along the
axes illustrating their distributions. Local moments at the atom site $Z$ are
omitted as they are nonmagnetic. Because 73.4\% of the test compounds are
nonmagnetic, both histograms exhibit a pronounced peak at zero. For magnetic
systems, the global sign is adjusted so that the total magnetic moment is
positive; since most magnetic compounds are ferromagnetic, positive moments
dominate in the distribution. Nearly all points fall along the diagonal and only
1.4\% of points lie along either axis, demonstrating that the model accurately
predicts both the magnitude and sign of local moments, and reliably
distinguishes ferromagnetic (FM) and ferrimagnetic (FiM) systems. For the
magnetic/nonmagnetic classification, we evaluated performance using receiver
operating characteristic (ROC) and precision-recall (PR) curves, as shown in
Fig.~S6. The area under the curve (AUC) values are 0.98 and 0.97, respectively,
indicating highly accurate classification.

Predicting \gls{lmm} is a common yet computationally demanding step in \gls{htp}
studies of magnetic materials. The approach developed in this work achieves high
accuracy in \gls{lmm} prediction and is readily transferable to other systems. A
central question, however, is how many training compounds are required to reach
satisfactory accuracy. To address this, we performed a learning-curve analysis
by training the \gls{mlrm} on progressively larger subsets of the dataset and
evaluating performance on a fixed test set of 5000 compounds, of which
1486 are magnetic. For evaluation, local moments at the $X_1$, $X_2$, and $Y$
sites were concatenated across samples into a single array, while the $Z$ site
was excluded since it is nonmagnetic in conventional Heusler compounds.

The learning curve is shown in Fig.~\ref{fig:MLRM_mag} (b), illustrating how model
performance improves with increasing training set size. In the top panel, the
$R^2$ scores for both local moments and their magnitudes are presented. The gap
between the two curves indicates that, while the model generally captures the
magnitude accurately, it sometimes assigns the incorrect sign. For example,
Mn$_2$ScGe with DFT-computed local moments \{$2.62$, $3.03$, $-0.29$,
$-0.10$\}\,$\mu_\mathrm{B}$ is predicted as \{$-2.71$, $3.04$, $0.02$,
$-0.02$\}\,$\mu_\mathrm{B}$ by \gls{mlrm}. The bottom panel reports two key
metrics: (i) classification accuracy for identifying magnetic systems and (ii)
the fraction of local moments with absolute error below 0.1\,$\mu_\mathrm{B}$,
evaluated across all compounds and within the magnetic subset. With 5000
training samples, the model achieves a classification accuracy of 0.92, and 90\%
of all compounds fall within the 0.1\,$\mu_\mathrm{B}$ error threshold. However,
this fraction decreases to 72\% when restricted to magnetic compounds subset,
indicating that the model identifies whether a site is magnetic with high
reliability but remains less accurate in predicting exact \gls{lmm} values.
Increasing the training set to 125,000 samples improves this fraction to 82\%,
while relaxing the threshold to 0.2\,$\mu_\mathrm{B}$ further raises it to 92\%.
Although performance improves with larger datasets, the gains become
progressively smaller. These results highlight the critical role of dataset size
in improving \gls{lmm} accuracy and inform the selection of training size in
future work focusing on other magnetic systems.

\subsection{Frozen transfer learning for \gls{mlrm} construction \label{sec:tlmlrm}}

To improve the performance of the \gls{mlrm}, we employed a frozen transfer
learning strategy using the \gls{esen} \gls{mlip} as the base model. The
\gls{esen} \gls{mlip} was trained on the OMat, MPtrj, and sAlexandria databases,
providing a comprehensive training set spanning the periodic table
\cite{barroso-luqueOpenMaterials20242024, jainCommentaryMaterialsProject2013,
dengCHGNetPretrainedUniversal2023,
schmidtMachineLearningAssistedDeterminationGlobal2023,
schmidtImprovingMachinelearningModels2024}. Through this training, the embedding
and the first several layers learn general chemical and structure
representations. To leverage this, we transferred the parameters of the
embedding and the first several layers into our \gls{mlrm} and kept them fixed
(frozen layers), while only updating the remaining layers and the output layer
(flexible layers). This approach is analogous to a recent work in which the
initial layers of ORB, EqV2, or MACE \glspl{mlip} were used to generate feature
vectors that were subsequently passed to property prediction models such as
MODNet, XGBoost, and MLP \cite{kimLeveragingNeuralNetwork2025a}.

\begin{figure}[htbp]
	\centering
	\includegraphics[width=0.95 \textwidth]{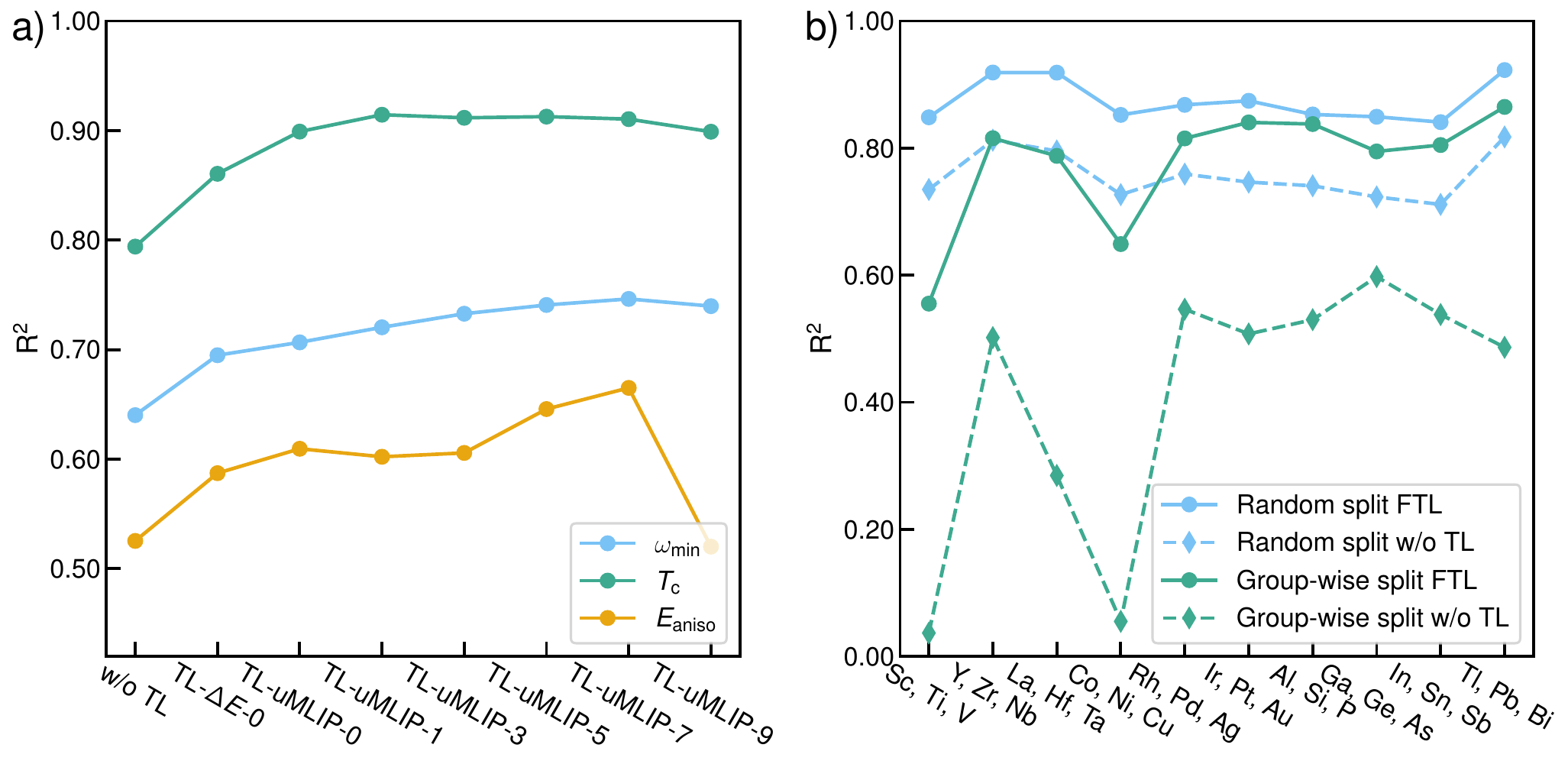}
	\caption{\label{fig:MLRM_FTL}
	    Frozen transfer learning performance and generalization.
		a) $R^2$ of models initialized from the
		\gls{esen} \gls{mlip} and trained with different numbers of frozen
		layers (denoted TL-uMLIP-$n$). In the $n=0$ case, the embedding layer is
		also trainable. A model trained from scratch (denoted w/o TL) and a
		transfer learning variant initialized from a base model pre-trained on
		\gls{form} data from the \gls{dxmag} database (denoted TL-$\Delta E$-0)
		are included for comparison. 
		b) Comparison of model performance on
		group-wise splits versus random splits. Models were evaluated on
		predicting \gls{tc}. Results are shown for frozen transfer learning
		(FTL) and models trained from scratch (w/o TL). For each test, three
		neighboring elements within the same period were used as the holdout
		elements. Six sets of d-block elements and four sets of p-block elements
		were tested.}
\end{figure}

The \gls{esen} model consists of 10 layers. Fig.~\ref{fig:MLRM_FTL} (a) shows the
$R^2$ scores of models trained with different numbers of frozen layers, denoted
TL-uMLIP-$n$. In the $n=0$ case, the embedding layer is also flexible. Results
for \gls{mpf}, \gls{tc}, and \gls{mae} are presented. Model performance improves
as the number of frozen layers increases up to $n=7$, after which it declines
when more layers are frozen. This trend reflects the balance between
transferring knowledge from the base model and maintaining sufficient
flexibility to adapt to the new task. Based on this analysis, TL-uMLIP-7 was
used in the ML-HTP case study for \gls{mpf}, \gls{tc}, \gls{mae}, and \gls{lmm}
\gls{mlrm}. For comparison, a model trained from scratch (w/o TL) is included,
which yields lower $R^2$ scores and highlights the benefit of transfer learning.
We also tested a variant initialized from a base model pre-trained on formation
energy ($\Delta E$) data from the \gls{dxmag} database (denoted TL-$\Delta
E$-0). TL-uMLIP-0 outperforms TL-$\Delta E$-0, indicating that initialization
from the \gls{esen} model, trained on a comprehensive dataset, is more
advantageous.

Besides improving overall performance, transfer learning from \gls{mlip} can
significantly enhance model generalization to unseen elements. To demonstrate
this, we evaluated model performance using group-wise split analysis. For each
group-wise split test, three neighboring elements within the same period were
used as the holdout elements. Any compound containing at least one of the
holdout elements was reserved as the holdout test data; the remaining compounds
were split into train/val with a 9:1 ratio.  This setup simulates realistic
scenarios where models are applied to material systems containing elements not
present in the training data, a typical domain shift in materials science. For
direct comparison, we also measured performance using train/val/test sets
created using random splits while keeping the counts of train/val/test the same
as in the group-wise split. We tested six sets of d-block elements and four sets
of p-block elements as holdout sets. We used \gls{tc} as the target property for
efficiency. Results obtained using frozen transfer learning and models trained from
scratch are shown in Fig.~\ref{fig:MLRM_FTL} (b).

Using FTL, performance on group-wise splits is generally lower than on random
splits, indicating that the model performance on unseen elements is reduced. The
drop in performance for p-block holdouts is smaller, which is consistent with
the fact that p-block elements are typically nonmagnetic in these compounds and
thus less directly related to the target property \gls{tc}. For group-wise
split tests with models trained from scratch, a much larger performance decrease
is observed. This occurs because FTL transfers the general chemical
representation learned by \gls{mlip} to property regressors, whereas models trained
from scratch do not benefit from this prior knowledge. This highlights the
advantage of transfer learning from \gls{mlip} for improving generalization to unseen
elements. Variation in random-split performance is due to different train set
sizes since the number of holdout compounds varies with the chosen holdout
elements. 

It should be noted that \gls{tl} is not always effective. It can be limited when
the representations learned by base model do not generalize to the target
domain. For example, source and target may differ substantially in material
class (e.g. bulk versus surface or molecular systems), or downstream task
concerns different physics (e.g. static energetics versus dynamic properties)
\cite{yamadaPredictingMaterialsProperties2019a, chenGraphNetworksUniversal2019a,
changOvercomingDataScarcity2022}. Furthermore, a recent work, which transfer
from model trained using computational data to learn experimental data,  
revealed that the error of the transfer-learned model decreases according to a
power-law as the size of the computational data for the base model increases
\cite{minamiScalingLawSim2Real2025}. This suggests that transfer learning may
also be less effective when the base model is trained on a small dataset even if
the base model performance within its own domain is good.

\section{Discussion}

We demonstrated the feasibility of combining \gls{mlip}s and \gls{mlrm}s for
\gls{htp} screening. As a case study, we identified 334 conventional quaternary
Heusler compounds and 924 all-$d$ Heusler compounds that exhibit thermodynamic,
dynamical, and magnetic stability, together with large \gls{mae}. The precision
of this workflow was confirmed through DFT validation of the candidate list.

For other material systems, if database for a subset is available, the same
workflow can be applied to explore the remaining chemical space. For novel
materials, \gls{mlip} can be used directly to reduce the search space by
filtering for thermodynamic stability, while target properties need to be
computed for a representative subset of compounds to train \glspl{mlrm} for
ML-HTP screening of the broader space. 

The \gls{mlrm}s in this study involve a domain shift, as the 
train dataset does not contain quaternary or all-$d$ Heusler compounds, yet the
models are applied to these systems. While performance is satisfactory, further
gains are possible using active learning approach. Iteratively refining the
models by selecting informative compounds can improve performance 
and help identify candidates missed by the current one-shot approach. In
such iterative framework, current DFT validation results can be added to the training
set, forming the first iteration of model refinement. This iterative strategy
naturally extends the present work, enabling more thorough 
exploration of chemical space while progressively improving model accuracy. Such
an approach is especially valuable when the target material space differs
substantially from the training data.

In the framework of this study, \gls{mlip} perform the critical task of
structure optimization, which is traditionally handled by DFT-derived methods,
while \glspl{mlrm} predict the properties of the optimized structures, tasks
that are typically carried out using DFT or DFT-derived methods. Together,
\gls{mlip} and \glspl{mlrm} enable a drop-in replacement for DFT in conventional
HTP workflows. This replacement is general and can be readily extended to other
properties, material classes, and DFT-based HTP pipelines, enabling accelerated
HTP screening and discovery of new materials.

\section{Methods}\label{sec:methods}

\subsection{ML-HTP workflow}

The ML-HTP workflow is schematically illustrated in Fig.~\ref{fig:schematic}(b)
and described in detail below.

In step (a) of ML-HTP workflow, the structures were optimized and the
\gls{form} and \gls{hull} were calculated using uMLIP.
The initial lattice constant was estimated as the average value of known $X_2YZ$
Heusler compounds in the \gls{dxmag} database that share two elemental species
with the target composition. A conventional cell in cubic phase with this
estimated lattice constant was then constructed. To generate initial structures,
the lattice parameters of this cell were systematically varied: the $a$, $b$,
and $c$ were uniformly scaled by $\pm10\%$ or $\pm30\%$, and, alternatively, the
$c$ alone was scaled by $\pm10\%$ or $\pm30\%$. All generated structures were
subsequently converted to the primitive unit cell and relaxed using the
\gls{esen} \gls{mlip}. The structure with the lowest energy after relaxation was
identified as the ground state. The choice of \gls{esen} was motivated by its
superior performance relative to other \glspl{mlip}, as discussed in
Sec.~\ref{sec:mlip_opti}. The selection of initial structures was validated by a
convergence test, which is provided in Fig.~S4.

For a compound to be thermodynamically stable against decomposition into its
constituent elements or competing phases, the formation energy must be negative
($\Delta E < 0$) and the distance to the convex hull must be zero ($\Delta H =
0$). In practice, however, metastable phases ($\Delta H > 0$) at 0 K may become
stabilized at finite temperature \cite{sunThermodynamicScaleInorganic2016}.
Following our previous work, we adopt a practical stability criterion of $\Delta
E < 0.0$~eV/atom and $\Delta H < 0.22$~eV/atom, which has been shown to
effectively capture experimentally accessible compounds
\cite{xiaoHighthroughputComputationalScreening2025}. Using the energies of
ground state candidates, constituent elements and competing phases predicted by
\gls{esen}, we computed $\Delta E$ and $\Delta H$ to assess thermodynamic
stability.

In step (b) of workflow, 
the \gls{mpf}, \gls{lmm}, \gls{tc}, and
\gls{mae} were predicted with \glspl{mlrm} trained on the \gls{dxmag} and
additional computed data. 
The \gls{mlip}-optimized structures were used as inputs.
The construction of these \glspl{mlrm} is described in
Sec.~\ref{sec:moment} and Sec.~\ref{sec:tlmlrm}. Compounds were classified as
dynamically stable for \gls{mpf} $>-10$~cm$^{-1}$, magnetic for $\sum_{i}
|\bm{m}_{i}| > 0.1~\mu_\mathrm{B}$/f.u., magnetically stable for \gls{tc} $>
300$~K, and strong \gls{mae} candidates for $|$\gls{mae}$| > 1$~MJ/m$^3$.
Tetragonal compounds satisfying all of these conditions were designated as
promising stable materials with strong \gls{mae}.

In step (c) of workflow, 
the candidate list was validated with \gls{dft} calculations to assess the
reliability of the \gls{ml}-\gls{htp} workflow. Structure optimizations were
performed starting from various initial spin configurations, consistent with our
previous DFT-HTP work. For conventional Heusler compounds, the magnetic moments
at the $X_1$, $X_2$, and $Y$ sites were initialized in configurations where they
were either parallel or antiparallel to each other. For all-$d$ Heuslers, spin
arrangements on all four sites were considered. To capture possible high-spin
and low-spin states, two initial magnitudes of the local moments ($|\bm{m}_{i}|
= 1$ and $4~\mu_\mathrm{B}$) were tested, along with a nonmagnetic
configuration ($|\bm{m}_{i}|=0$). The \gls{mlip}-optimized structures served as
the starting geometries. After structure relaxation, the ground state was
identified by comparing total energies. For the resulting ground states, we
computed $\Delta E$, $\Delta H$, \gls{mae}, phonon, and \gls{tc} using VASP,
OQMD, ALAMODE, and SPRKKR \cite{saalMaterialsDesignDiscovery2013,
bahnObjectorientedScriptingInterface2002,
tadanoSelfconsistentPhononCalculations2015, AnharmonicForceConstants,
ebertCalculatingCondensedMatter2011, liechtensteinLocalSpinDensity1987}.

\subsection{Computational methods}

The \gls{mlip}-based structural optimizations were performed using the \gls{ase}
package \cite{hjorthlarsenAtomicSimulationEnvironment2017}. The Fast Inertial
Relaxation Engine (FIRE) optimizer was employed, with symmetry constraints
enforced throughout the relaxation process
\cite{bitzekStructuralRelaxationMade2006}. To ensure consistency in reference
energies for computing $\Delta E$ and $\Delta H$, the elemental phases and
competing phases were also optimized using the same \gls{mlip}. Their initial
geometries were taken from DFT-optimized structures in the \gls{oqmd} database
\cite{saalMaterialsDesignDiscovery2013,
bahnObjectorientedScriptingInterface2002}.

The \glspl{mlrm} were developed to predict \gls{mpf}, \gls{lmm}, \gls{tc}, and
\gls{mae}. To leverage prior knowledge, we employed a frozen transfer learning
strategy, as illustrated in Fig.~\ref{fig:schematic}(a). Each \gls{mlrm} was
initialized from the pretrained \gls{esen} \gls{mlip}, with the embedding layer
and the first seven message-passing layers kept frozen, while the final three
layers and the output layer were fine-tuned. This framework was implemented
using a modified version of the FAIRChem package
(v1)~\cite{shuaibiFAIRChem2025}. For the \gls{lmm} \gls{mlrm} training, the loss
function was adapted to address the global sign ambiguity of magnetic moments,
as described in Sec.~\ref{sec:moment}. The training dataset consisted of
\gls{dxmag} together with newly computed \gls{mae} and \gls{tc} values based on
the optimized structures in \gls{dxmag}. The data were randomly partitioned into
training, validation, and test sets in an 8:1:1 ratio. For \gls{lmm}, we used
all ground-state entries, resulting in 27,864 data points. The \gls{mpf} data
were available for all thermodynamically stable ground states, yielding 8,198
entries. For \gls{tc}, 2,106 data points were used, including 750 newly computed
values. Since \gls{mae} data were not included in \gls{dxmag}, we calculated
\gls{mae} for all magnetic tetragonal ground-state systems, obtaining 6,123
entries.

\Gls{dft} calculations were performed primarily with
\gls{vasp}~\cite{kresseEfficiencyInitioTotal1996,
kresseEfficientIterativeSchemes1996}, using the projector augmented wave (PAW)
method and the generalized gradient approximation (GGA) with the
Perdew–Burke–Ernzerhof (PBE)
functional~\cite{kresseUltrasoftPseudopotentialsProjector1999,
perdewGeneralizedGradientApproximation1996}. \gls{form}, \gls{hull}, phonon, and
\gls{tc} were computed using OQMD, ALAMODE, and SPRKKR following the methodology
of our previous DFT-HTP study of ternary Heusler
compounds~\cite{saalMaterialsDesignDiscovery2013,
bahnObjectorientedScriptingInterface2002,
tadanoSelfconsistentPhononCalculations2015, AnharmonicForceConstants,
ebertCalculatingCondensedMatter2011, liechtensteinLocalSpinDensity1987,
xiaoHighthroughputComputationalScreening2025}. 
The agreement of magnetic moments between \gls{vasp} and SPRKKR is demonstrated in Fig.~S7.
Phonon calculations for the
all-$d$ compounds MnOsMnRe and MnReMnRu failed due to convergence issues in DFT,
and these compounds were treated as unstable in the validation rate
analysis.
\gls{mae} was calculated as
\gls{mae}$=E_{\perp}-E_{\|}$ using the force theorem
\cite{daalderopFirstprinciplesCalculationMagnetocrystalline1990,
xingLatticeDynamicsIts2022, xingFirstprinciplesPredictionPhase2023}.
Calculations were performed in the primitive cell with $\bm{k}$-meshes generated
using \gls{pmg} at a density of \SI{6000}{\angstrom^{-3}}, and the tetrahedron
method with Bl\"ochl corrections was
applied~\cite{ongPythonMaterialsGenomics2013,
blochlImprovedTetrahedronMethod1994}. Input generation, structure manipulation,
and symmetry analysis were carried out using \gls{pmg}, \gls{ase}, ASE2SPRKKR,
and spglib~\cite{ongPythonMaterialsGenomics2013,
hjorthlarsenAtomicSimulationEnvironment2017, ASE2SPRKKRSoftwarePackage,
bahnObjectorientedScriptingInterface2002, togoSpglibSoftwareLibrary2024}.

\subsection{Computational cost}

The computational cost for each task in the HTP workflow is summarized in
Table~\ref{tab:efficiency_comparison}. DFT timings are based on validation runs
for ML-selected candidates, while ML timings are taken from the ML-HTP
screening. For DFT, we report per-job statistics such as mean, median, and
interquartile range (IQR). \gls{mlip} structure optimization and MLRM predictions
were performed in batches, and mean values are reported as
individual per-job timings are not available. MLRM training times correspond to
the wall time required to train a single model on one GPU. DFT calculations
except phonon were performed on dual-socket Intel Xeon Platinum 8268 (Cascade
Lake, 24 cores per CPU, 2.9 GHz, 48 cores per node), phonon calculations were
performed on Fujitsu A64FX processors (Armv8.2-A SVE 512 bit, 48 compute cores
per CPU, 2.2 GHz), \gls{mlip} structure optimizations were performed on dual-socket
Intel Xeon Gold 6230 (Cascade Lake, 20 cores per CPU, 2.1 GHz, 40 cores per
node), and MLRMs were trained and applied on NVIDIA RTX 6000 Ada GPUs. All DFT
calculations, except for phonon calculations, were performed on 2 nodes; phonon
calculations were performed on 6 nodes. The \gls{mlip} structure optimizations were
performed on 1 node and MLRM training and prediction were performed on 1 GPU. 

\begin{table}[h]
\centering
\renewcommand{\arraystretch}{1.2}
\begin{tabular}{@{}lccc@{}}
\toprule
 & \shortstack{DFT \\ (mean/median/IQR)} & \shortstack{ML training \\ (mean)} & \shortstack{ML prediction \\ (mean)} \\
\midrule
		Structure relaxation & 5.6/3.8/3.0     & --- & 0.12     \\
		\gls{lmm}            & ---             & 876 & $<$1e-03 \\
		Phonon stability     & 4542/4008/1442  & 258 & $<$1e-03 \\
		\gls{tc}             & 70/66/14        & 78  & $<$1e-03 \\
		\gls{mae}            & 142/133/34      & 192 & $<$1e-03 \\
\bottomrule
\end{tabular}
\caption{Computational cost of each task in the HTP workflow, reported in
node-minutes. DFT entries show per-job statistics (mean/median/IQR). ML entries
show mean values. For structure relaxation, one job corresponds to a single
relaxation from an initial distortion. For structure optimization, the cost per
compound should be multiplied by the number of initial structures, which varies
across material systems; here, we report the cost for a single structure
relaxation. The \gls{mlip} is pre-trained and used without fine-tuning, so its
training time is excluded. DFT structure optimizations directly yield local
magnetic moments, so no separate timing is reported.}
\label{tab:efficiency_comparison}
\end{table}

In the ML‑HTP case study, \gls{mlip} structure optimization, MLRM training, and MLRM
prediction used 1,835 node‑hours, 23 GPU‑hours, and 4 GPU‑hours, respectively.
DFT calculations of ML‑selected candidates consumed 1,645 node‑hours for
optimizations and 99,680 node‑hours for property evaluations. Ignoring the node
difference, the total ML‑HTP cost is 103,160 node‑hours
plus 27 GPU‑hours. A DFT-HTP workflow screening the same chemical
space would need about 256,160 node‑hours for structure optimizations and
18,802,625 node‑hours for property evaluations, estimated using statistics
reported in Table~\ref{tab:efficiency_comparison}. This comparison shows an
estimated speed‑up of 185 times for ML‑HTP with DFT validation, or 10$^4$ times
if used without DFT validation at the end. Please note that these estimates are
approximate, and waiting time for job execution, human time spent on debugging
and workflow management are not included. The actual speed-ups also vary based
on material systems, target properties, and computational resources.

\section*{Data availability}
The ML-HTP candidate list and DFT validation results are included in the
Supplementary Information as Tables S3 and S4. The complete set of all screened compounds, along with
ML-predicted properties, will be made available through the \gls{dxmag} database
at \url{https://www.nims.go.jp/group/spintheory/}.

\section*{Code availability}
The developed packages \gls{hot} and \gls{ftl} will be made available through
the Spin Theory Group GitHub repository at
\url{https://github.com/nims-spin-theory} and our group website at
\url{https://www.nims.go.jp/group/spintheory/}.

\section*{Funding}
This study was supported by MEXT Program: Data Creation and Utilization-Type
Material Research and Development Project (Digital Transformation Initiative
Center for Magnetic Materials) Grant Number JPMXP1122715503 and as ``Program for
Promoting Researches on the Supercomputer Fugaku'' (Data-Driven Research Methods
Development and Materials Innovation Led by Computational Materials Science,
JPMXP1020230327). 

\section*{Acknowledgments}
This study used computational resources of supercomputer
Fugaku provided by the RIKEN Center for Computational Science (Project ID:
hp240223), the computer resources provided by ISSP, U-Tokyo under the program of
SCCMS, and the computer resources at NIMS Numerical Materials Simulator.

\section*{Author contributions}
TT conceptualized, designed, and supervised the project; reviewed and edited
the manuscript. TT and EX developed the methodology and code implementation;
performed the calculations and analysis; EX drafted the manuscript.

\section*{Competing interests}
All authors declare no financial or non-financial competing interests.


\bibliography{ref}

\end{document}